\documentclass{article}
\usepackage{amsmath, amsthm, amssymb}
\usepackage[left=1in, bottom=1in, right=1in, top=1in]{geometry}
\usepackage{xcolor}
\usepackage[comma,authoryear,round]{natbib}
\usepackage{soul}
\usepackage{graphicx}
\usepackage{dsfont}

\usepackage{float}
\usepackage{bbm}
\usepackage{authblk}
\usepackage{comment}
\usepackage{longtable}
\usepackage[font=normalsize]{caption}   
\usepackage[title]{appendix}
\usepackage{csquotes}
\usepackage{caption}
\usepackage{lipsum}

\title{Modeling Multivariate Positive-Valued Time Series Using R-INLA}

\author[a]{Chiranjit Dutta \thanks{Corresponding Author: chiranjit.dutta@uconn.edu}}
\author[b]{Nalini Ravishanker}
\author[c]{Sumanta Basu}

\affil[a,b]{Department of Statistics, University of Connecticut, Storrs, CT, USA}
\affil[c]{Department of Statistics and Data Science, Cornell University, Ithaca, NY, USA}

\date{\today}

\begin{document}
\maketitle

\begin{abstract}

In this paper we describe fast Bayesian statistical analysis of vector positive-valued time series, with application to interesting financial data streams. We discuss a flexible level correlated model (LCM) framework for building hierarchical models for vector positive-valued time series. The LCM allows us to combine marginal gamma distributions for the positive-valued component responses, while accounting for association among the components at a latent level. We use integrated nested Laplace approximation (INLA) for fast approximate Bayesian modeling via the \texttt{R-INLA} package, building custom functions to handle this setup. We use the proposed method to model interdependencies between realized volatility measures from several stock indexes.

\end{abstract}


\section{Introduction} \label{intro}

Multivariate positive-valued time series are ubiquitous across diverse application domains, including but not limited to, epidemiology, econometrics, finance, insurance, and signal processing. 
In financial applications, we increasingly observe long multivariate positive-valued time series from multiple stocks. Examples include time series related to high-frequency trading, such as volume, price, realized volatility measures, durations, etc. While there has been a considerable amount of research on modeling univariate positive-valued time series with exponential and gamma marginals (\cite{lawrance1980exponential,lawrance1981new,lawrance1985modelling}, \cite{jacobs1977mixed}, \cite{raftery1982generalized}, \cite{gaver1980first} and \cite{gourieroux2006autoregressive}),
research on the multivariate setup is rather limited. 
A few references which deal with modeling multivariate positive-valued time series include \cite{prekopa1978new}, \cite{yue2001bivariate}, \cite{cipollini2006vector} and \cite{hautsch2013modeling}.
While these \textit{static} modeling approaches may be useful when the time series are stationary, many interesting situations in practice involve nonstationary time series and will benefit from fitting \textit{dynamic} models.

Gaussian dynamic models have been widely used for several decades for modeling univariate and multivariate time series \citep{west2006bayesian,petris2010r}. 
More recently, dynamic generalized linear models (DGLM's) 
have been tailored to model non-Gaussian time series without the necessity of
data transformations, or other adaptations; see
\cite{west1985dynamic}, 
\cite{grunwald1993prediction}, \cite{fahrmeir1987regression}, \cite{lindsey1995dynamic}, \cite{gamerman1998markov}, \cite{chiogna2002dynamic}, \cite{shephard1997likelihood}, \cite{benjamin2003generalized} and \cite{godolphin2006decomposition}.  
\cite{fahrmeir1994multivariate} and \cite{durbin2012time} are excellent books on this topic.

Non-Gaussian dynamic modeling with marginals having positive support (such as gamma, lognormal, or Weibull)  is an attractive setup for analyzing multivariate positive-valued  series exhibiting temporal nonstationarity, as well as skewness and heavy-tails.
Univariate gamma regression models for each of the components of the response vector cannot account for the dependence among the components, where the dependence may be due to omitted variables which simultaneously effect the response vector \citep{aitchison1989multivariate,chib2001markov,ma2008multivariate}. Therefore, an adequate multivariate model for positive time series is needed.

To our knowledge, \cite{aktekin2020family} is the only paper describing a dynamic model for multivariate positive-valued time series. It accounts for correlations among the components through a common random environment that evolves over time. They develop MCMC methods as well as sequential Monte Carlo methods for parameter estimation.
\cite{tsionas2004bayesian} described  Bayesian analysis using a 
multivariate gamma likelihood \citep{mathai1991multivariate}, non-conjugate prior specifications, and  
Gibbs sampler with data augmentation. 
This is not a dynamic model, and moreover, computing the  
likelihood function can be cumbersome and time consuming. Incorporating the dependence among the components of the positive-valued vector through a different mechanism will be useful.
{In general, using Markov chain Monte Carlo (MCMC) methods for Bayesian hierarchical dynamic modeling of vector positive-valued time series under a multivariate gamma sampling distributional assumption may be computationally demanding, especially since there does not exist any convenient form of the likelihood function. 

The level correlated model (LCM) framework described in \cite{serhiyenko2015dynamic} for multivariate time series of counts offers a computationally simpler way of handling dependency among the components in vector-valued responses.
We adapt this framework for modeling multivariate positive-valued time series, which assumes marginal gamma component distributions, and accounts for the association between the components at a latent Gaussian level. This avoids the need for computing the multivariate gamma likelihood, which can be computationally demanding.}

Another way to reduce the computational burden is to replace the MCMC approach by  the integrated nested Laplace approximation (INLA) approach.  
INLA enables approximate Bayesian inference for a large class of latent Gaussian models (LGMs) with wide-ranging applications \cite{rue2009approximate}, 
and unlike simulation intensive Bayesian approaches, it performs approximate inference using a series of deterministic approximations that take advantage of the LGM structure to provide fast and accurate inference. 
\cite{ruiz2012direct} presented a general framework which enabled users to use \texttt{R-INLA} for a variety of state-space models.  
\cite{serhiyenko2015approximate,serhiyenko2018multi} used multivariate 
Gaussian random effects to model level correlated effect in multivariate time series of counts in 
the LCM framework thus facilitating the estimation of LCMs using \texttt{R-INLA} package.

In this article, we describe the LCM framework for 
dynamic modeling of vector  positive-valued time series. 
At the observation level, the components are conditionally independent with marginal gamma distributions, while a level correlation random effect knits the dependence between components.
The latent state vector evolves as a stationary vector autoregression (VAR) or vector autoregression and moving average (VARMA) process. 
We use \texttt{R-INLA} for the data analysis, developing custom code via the \texttt{rgeneric()} function for implementing the 
VAR or VARMA latent models.
This setup is useful for financial modeling of
the joint dynamics of realized volatility measures among the stock indexes. 

The format of the paper follows. Section \ref{LCM} describes the general framework of the level correlated model (LCM). Section \ref{Approx Bayes R-INLA} gives the description of the approximate Bayesian analysis using \texttt{R-INLA} including the derivation of the joint distribution of the latent states in subsection \ref{Joint Distribution}, followed by the prior specification and model fitting using \texttt{R-INLA} in subsection \ref{Prior}. For simulated data, Section \ref{Simulated Study} demonstrates that the \texttt{R-INLA} approach enables accurate model fitting. Section \ref{Empirical Results} presents an analysis of the multivariate realized measures of volatility. Finally, Section \ref{Discussion} concludes with discussion and directions for future research.

\section{Level Correlated Model Framework}\label{LCM}

Let $\mathbf{y}_{it} = (Y_{1,it},\dots, Y_{m,it})^{\prime}$ be a $m$-variate vector of continuous positive responses for $i=1,\dots,n$ and $t=1,\dots,T$. In the financial example, we would observe $m$ components of positive responses on $n$ stocks over $T$ days.
We describe the LCM setup for positive responses similar to the description in \cite{serhiyenko2015dynamic}. In the following, we let $i=1\dots n$, $t=1\dots T$, and $j=1 \dots m$.
Assuming gamma marginals for the components, the observation equation can be written as
\begin{equation}
\label{Eq: Gamma marginal}
    Y_{j,it}|\theta_{j,it},\tau \sim \text{Gamma}\bigg(\tau,\dfrac{\tau}{\theta_{j,it}}\bigg), \end{equation}
where 
the precision parameter $\tau >0$ is static, and the random mean $\theta_{j,it} >0$. 
Then, $\text{E}(Y_{j,it}|\theta_{j,it},\tau) = \theta_{j,it}$, and 
$\text{Var}(Y_{j,it}|\theta_{j,it},\tau) = \theta_{j,it}^2/\tau$. A dynamic model for the random mean $\theta_{j,it}$ is 
\begin{equation}
\label{Eq: Gamma Obs Eqn}
    \log(\theta_{j,it}) = \beta_{j,i0} + x_{j,t} +  \boldsymbol{S}^{\prime}_{j,it} \boldsymbol{\beta}_{j,i} + \alpha_{j,it},
\end{equation}
where 
the random effect $\beta_{j,i0}$ is a subject-specific intercept 
corresponding to the $j$th component of the response, $x_{j,t}$ is a time-varying latent state vector which depends on the $j$th component,
and  the random effect $\alpha_{j,it}$ is a response type, time, and subject-specific level correlated error component. The vector $\boldsymbol{S}_{j,it}^{\prime}$ denotes a $p_{j}$-dimensional vector of covariates, and $\boldsymbol{\beta}_{j,i}$ is a corresponding $p_{j}$-dimensional vector of coefficients. In general, $\boldsymbol{\beta}_{j,i}$ can also be treated as time-varying. 
We have chosen to use the logarithmic link function in (\ref{Eq: Gamma Obs Eqn}) corresponding to the gamma marginals in (\ref{Eq: Gamma marginal}), but other suitable link functions may also be used.
Let the level correlation vector $\boldsymbol{\alpha}_{it} = (\alpha_{1,it},\dots,\alpha_{m,it})^{\prime}$.
The dependence between the components of $\boldsymbol{y}_{it}$ can be introduced 
via 
$\boldsymbol{\alpha}_{it} \sim N(\boldsymbol{0},\boldsymbol{\Sigma})$, where $\boldsymbol{\Sigma}$ is a is a variance-covariance matrix for the level correlated error.  
More generally, we can allow $\boldsymbol{\Sigma}$ to be 
a subject-specific parameter and assume that  $\boldsymbol{\alpha}_{it} \sim N(\boldsymbol{0},\boldsymbol{\Sigma}_{i})$. 

We specify the evolution of the latent state vector $\boldsymbol{x}_t = (x_{1,t},\dots,x_{m,t})^{\prime}$ via a vector autoregression, VAR$(p)$, given by 
\begin{equation}\label{Eq: VAR_state}
    \boldsymbol{x}_t = \sum_{h=1}^{p}\boldsymbol{\Phi}_h \boldsymbol{x}_{t-h} + \boldsymbol{w}_t =
    \boldsymbol{\Phi}(B)\boldsymbol{x}_t +
    \boldsymbol{w}_t, 
\end{equation}
where $t=1\dots T$, $\boldsymbol{\Phi}_{h}$ are coefficient matrices of order $m \times m$ and the error vector $\boldsymbol{w}_t \sim \text{N}(\boldsymbol{0},\boldsymbol{W})$, $\boldsymbol{W}$ being a $m\times m$ state error variance-covariance matrix. 
In \eqref{Eq: VAR_state}, 
$\boldsymbol{\Phi}(B)$ is the 
VAR($p$) operator defined as 
\begin{equation}\label{Eq: VAR_operator}
    \boldsymbol{\Phi}(B) = \boldsymbol{I}_{m} - \boldsymbol{\Phi}_{1}B - \dots - \boldsymbol{\Phi}_{p}B^{p},
\end{equation}
where $B$ is the back-shift operator such that $B^{j} \boldsymbol{x}_{t} = \boldsymbol{x}_{t-j}$, for any $j \geq 1$.
The VAR($p$) process is stationary if the roots of the determinant of $\boldsymbol{\Phi}(z)$ are outside the unit circle, i.e., the modulus of each root is greater than 1, \citep{lutkepohl2005new}. We refer to the model defined by (\ref{Eq: Gamma marginal})-
(\ref{Eq: VAR_state}) as an LCM-VAR, with independent gamma marginals at the observation level. Let $\boldsymbol{\Theta}^{*}= (\beta_{j,i0},\boldsymbol{\beta}^{\prime}_{j,i},\text{vec}^{\prime}(\boldsymbol{\Sigma}),\text{vec}^{\prime}(\boldsymbol{\Phi}_{1}),\dots,\text{vec}^{\prime}(\boldsymbol{\Phi}_{p}),\text{vec}^{\prime}(\boldsymbol{W}))^{\prime}$ denote the set of all LCM-VAR parameters, where $\text{vec}$ is the vectorization operator.

If the $\boldsymbol{\Phi}_h$ matrices in (\ref{Eq: VAR_state}) are all assumed to be diagonal, the LCM-VAR model simplifies to $m$ separate LCM-AR models, again with independent gamma marginals.
Here, each component of the latent state vector $\boldsymbol{x}_t$ evolves according to an autoregressive (AR) process of order $p$, and the state equation \eqref{Eq: VAR_state} becomes  
\begin{equation}
\label{AR state eq}
    x_{j,t} = \sum_{h=1}^{p}\phi_{j,h} x_{j,(t-h)} + w_{j,t},~j=1,\ldots,m,
\end{equation}
where 
the error $w_{j,t} \sim \text{N}(0,W_{j})$, for scalar $W_j$.

\section{Approximate Bayesian Analysis using INLA}\label{Approx Bayes R-INLA}

In the LCM framework, it is possible to use \texttt{R-INLA} templates when the state equation follows 
\eqref{AR state eq}, but not when it follows the VAR($p$) model in \eqref{Eq: VAR_state}.
To implement a model with latent VAR effects, we can use the \texttt{rgeneric()} function to 
code the joint p.d.f. of the state vector $\boldsymbol{x}_t$ which is  described in the next section. 



\subsection{Joint Distribution of the Latent States}\label{Joint Distribution}
Suppose  $\boldsymbol{x}_t = (X_{t,1},\dots,X_{t,m})^{\prime}$ is an $m$-dimensional latent process in
\eqref{Eq: VAR_state}.
Let $\boldsymbol{\Theta}$ denote the vector of all the parameters. 
The latent Gaussian Markov field representation is 
that of a multivariate Gaussian vector with a sparse precision matrix as in \citep{rue2005gaussian,JSSv098i02};  
i.e.,
\begin{equation}
    vec(\boldsymbol{\Theta}) \sim N(\boldsymbol{0},\boldsymbol{\Sigma}^{*}),
\end{equation}
where $\boldsymbol{0}$ denotes vector of zeros and $\boldsymbol{\Sigma}^{*}$ is the variance-covariance matrix.
Since \texttt{R-INLA} works with the precision matrix,
we derive 
the form of the joint p.d.f. in terms of  $\boldsymbol{\Sigma}^{*-1}$. We first show the form of the joint p.d.f. for the VAR(1) latent process.

\subsubsection{Joint p.d.f. for a VAR(1) Process}\label{sec:pdfvar1}

When $p=1$, \eqref{Eq: VAR_state} simplifies to a VAR(1) process:
\begin{equation}  \label{Eq: VAR1}
    \boldsymbol{x}_t = \boldsymbol{\Phi} \boldsymbol{x}_{t-1} + \boldsymbol{w}_t.
\end{equation}
To derive the joint distribution of $(\boldsymbol{x}_{1}^{\prime},\dots,\boldsymbol{x}_{T}^{\prime})^{\prime}$, we see that  
\begin{equation*}
    \log \pi(\boldsymbol{x}_t | \boldsymbol{x}_{t-1}) = - \frac{1}{2} \boldsymbol{w}_{t}^{\prime} \boldsymbol{W}^{-1} \boldsymbol{w}_{t} + C,
\end{equation*}
where $\boldsymbol{w}_t = \boldsymbol{x}_t - \boldsymbol{\Phi} \boldsymbol{x}_{t-1}$, $\pi(.)$ denotes the probability density function and C is a constant. We assume diffuse initial conditions for $\boldsymbol{x}_{1} \sim N(\boldsymbol{0},\frac{1}{\kappa}\boldsymbol{I}_{m})$, where $\kappa = 0.01$. 
The joint distribution of $(\boldsymbol{x}_{1}^{\prime},\dots,\boldsymbol{x}_{T}^{\prime})^{\prime}$ is given by
\begin{equation*}
\begin{aligned}
    &\log \pi(\boldsymbol{x}_1,\boldsymbol{x}_2,\dots\boldsymbol{x}_n) =  \sum_{t=2}^{n} \log \pi(\boldsymbol{x}_{t}|\boldsymbol{x}_{t-1}) +  \log \pi(\boldsymbol{x}_{1}) \\
    & = - \dfrac{1}{2}\big[(\boldsymbol{x}_{n} - \boldsymbol{\Phi} \boldsymbol{x}_{n-1})^{\prime} \boldsymbol{W}^{-1} (\boldsymbol{x}_{n} - \boldsymbol{\Phi} \boldsymbol{x}_{n-1}) + (\boldsymbol{x}_{n-1} - \boldsymbol{\Phi} \boldsymbol{x}_{n-2})^{\prime} \boldsymbol{W}^{-1} (\boldsymbol{x}_{n-1} - \boldsymbol{\Phi} \boldsymbol{x}_{n-2})\\
    & \hspace{3cm} + \dots + (\boldsymbol{x}_{2} - \boldsymbol{\Phi} \boldsymbol{x}_{1})^{\prime} \boldsymbol{W}^{-1} (\boldsymbol{x}_{2} - \boldsymbol{\Phi} \boldsymbol{x}_{1}) +  \boldsymbol{x}_{1}^{\prime} \kappa \boldsymbol{I}_{m} \boldsymbol{x}_{1}\big] + C'\\
    & = -\dfrac{1}{2}\big[\boldsymbol{x}_{n}^{\prime} \boldsymbol{W}^{-1} \boldsymbol{x}_{n} - \boldsymbol{x}_{n}^{\prime} \boldsymbol{W}^{-1} \boldsymbol{\Phi} \boldsymbol{x}_{n-1} - \boldsymbol{x}_{n-1}^{\prime} \boldsymbol{\Phi}^{\prime} \boldsymbol{W}^{-1} \boldsymbol{x}_{n} + \boldsymbol{x}_{n-1}^{\prime} (\boldsymbol{\Phi}^{\prime} \boldsymbol{W}^{-1} + \boldsymbol{W}^{-1})  \boldsymbol{x}_{n-1} \\
    & \hspace{3cm} + \dots + \boldsymbol{x}_{2}^{\prime} (\boldsymbol{\Phi}^{\prime} \boldsymbol{W}^{-1} \boldsymbol{\Phi} \boldsymbol{x}_{2} + \boldsymbol{W}^{-1}) \boldsymbol{x}_{2} - \boldsymbol{x}_{2}^{\prime} \boldsymbol{W}^{-1} \boldsymbol{\Phi}^{\prime} \boldsymbol{x}_{1} - \boldsymbol{x}_{1}^{\prime} \boldsymbol{\Phi}^{\prime} \boldsymbol{W}^{-1}\boldsymbol{x}_{2} \\
    & \hspace{3cm} + \boldsymbol{x}_{1}^{\prime} (\boldsymbol{\Phi}^{\prime} \boldsymbol{W}^{-1} \boldsymbol{\Phi} + \kappa \boldsymbol{I}_{m})\boldsymbol{x}_{1}\big ] + C'
\end{aligned}
\end{equation*}
Let $\boldsymbol{A} = \boldsymbol{\Phi}^{\prime} \boldsymbol{W}^{-1} \boldsymbol{\Phi}$, $\boldsymbol{B} = - \boldsymbol{\Phi}^{\prime} \boldsymbol{W}^{-1}$ and $\boldsymbol{C} = \boldsymbol{W}^{-1}$ be the $m \times m$ matrices and let $\boldsymbol{0}_m$ be the $m\times m$ matrix of 0's. Hence the precision matrix for the joint distribution of $(\boldsymbol{x}_1^{\prime},\dots,\boldsymbol{x}_{T}^{\prime})^{\prime}$ is a $mT \times mT$ matrix given by
\begin{equation}\label{Eq: VAR1_precision}
\boldsymbol{\Sigma}^{*-1}=\begin{bmatrix}
\boldsymbol{A} + \kappa \boldsymbol{I}_{m}  & \boldsymbol{B}  & \cdots & \cdots & \cdots & \cdots & \cdots & \boldsymbol{0}_m \\
\boldsymbol{B}^{\prime}  & \boldsymbol{A+C}  & \boldsymbol{B}  & \ddots & && & \vdots \\
 \vdots  & \boldsymbol{B}^{\prime} & \boldsymbol{A+C}  & \boldsymbol{B} & &  & \vdots \\
\vdots & \ddots & \ddots & \ddots & \ddots & \ddots &  & \vdots \\
\vdots & & \ddots & \ddots & \ddots & \ddots & \ddots& \vdots\\
\vdots  & & & \ddots & \ddots  & \ddots  & \ddots  & \boldsymbol{0}_m\\
\vdots  & && & \ddots & \ddots  & \ddots  &  \boldsymbol{B}\\
\boldsymbol{0}_m & \cdots &  \cdots & \cdots & \cdots & \boldsymbol{0}_m & \boldsymbol{B}^{\prime}  & \boldsymbol{C}\\
\end{bmatrix}
\end{equation}

\subsubsection{Joint p.d.f. for a VAR($p$) Process}

Following \cite{lutkepohl2005new}, we can represent a VAR($p$) process of an $m$-dimensional time series  (in \eqref{Eq: VAR_state}) as a VAR(1) process of a $pm$-dimensional time series, by 
stacking $\boldsymbol{x}_{t},\boldsymbol{x}_{t-1},\dots \boldsymbol{x}_{t-p+1}$: 
\begin{eqnarray}
\begin{pmatrix}
\boldsymbol{x}_{t}\\
\boldsymbol{x}_{t-1} \\
\boldsymbol{x}_{t-2} \\
\vdots\\
\vdots\\
\boldsymbol{x}_{t-p+1} \\
\end{pmatrix}  &= & \begin{pmatrix}
\boldsymbol{\Phi}_{1} & \boldsymbol{\Phi}_{2} & \cdots & \cdots & \boldsymbol{\Phi}_{p-1} & \boldsymbol{\Phi}_{p}\\ \boldsymbol{I}_{m} & \boldsymbol{0}_{m} & \ddots &  & \boldsymbol{0}_{m} & \boldsymbol{0}_{m}\\
\boldsymbol{0}_{m} & \boldsymbol{I}_{m} & \ddots & \ddots & \boldsymbol{0}_{m} & \boldsymbol{0}_{m}\\
\vdots & \vdots & \ddots & \ddots & \ddots& \vdots\\
\boldsymbol{0}_{m} & \boldsymbol{0}_{m} & \cdots & \cdots & \boldsymbol{I}_m & \boldsymbol{0}_{m}\\
\end{pmatrix} \begin{pmatrix}
\boldsymbol{x}_{t-1}\\
\boldsymbol{x}_{t-2} \\
\boldsymbol{x}_{t-3} \\
\vdots\\
\vdots\\
\boldsymbol{x}_{t-p} \\
\end{pmatrix} + \begin{pmatrix}
\boldsymbol{w}_{t}\\
\boldsymbol{0} \\
\boldsymbol{0} \\
\vdots\\
\vdots\\
\boldsymbol{0} \\
\end{pmatrix}, \text{ i.e.,} \notag \\
\boldsymbol{x}^{*}_{t} &=& \boldsymbol{\Phi}^{*} \boldsymbol{x}^{*}_{t-1} + \boldsymbol{w}^{*}_{t},
\end{eqnarray}
where $\boldsymbol{w}^{*}_{t} \sim SN(\boldsymbol{0},\boldsymbol{W}^{*})$, $\boldsymbol{W}^{*} = \begin{pmatrix}
\boldsymbol{W} & \boldsymbol{0}_{m} & \cdots & \boldsymbol{0}_{m}\\ 
\vdots & \vdots & \ddots &  \vdots\\
\boldsymbol{0}_{m} & \boldsymbol{0}_{m} & \cdots & \boldsymbol{0}_{m}\\
\end{pmatrix}$ is a $pm \times pm$ p.s.d. matrix and $SN$ denotes a singular normal distribution.

Following the setup for the VAR$(1)$ process in Section 
\ref{sec:pdfvar1}, we can then obtain the $pmT \times pmT$ dimensional precision matrix for the joint distribution of $T$ realizations from a VAR($p$) process as 
\begin{equation}\label{VAR_p_precision}
\boldsymbol{\Sigma}^{*-1}=
\begin{bmatrix}
\boldsymbol{A}^{*} + \kappa^{*} \boldsymbol{I}_{m}  & \boldsymbol{B}^{*}  & \cdots & \cdots & \cdots & \cdots & \cdots & \boldsymbol{0}_m \\
\boldsymbol{B}^{*\prime}  & \boldsymbol{A}^{*} + \boldsymbol{C}^{*}  & \boldsymbol{B}^{*}  & \ddots & && & \vdots \\
 \vdots  & \boldsymbol{B}^{*\prime} & \boldsymbol{A}^{*} + \boldsymbol{C}^{*}  & \boldsymbol{B}^{*} & &  & \vdots \\
\vdots & \ddots & \ddots & \ddots & \ddots & \ddots &  & \vdots \\
\vdots & & \ddots & \ddots & \ddots & \ddots & \ddots& \vdots\\
\vdots  & & & \ddots & \ddots  & \ddots  & \ddots  & \boldsymbol{0}_m\\
\vdots  & && & \ddots & \ddots  & \ddots  &  \boldsymbol{B}^{*}\\
\boldsymbol{0}_m & \cdots &  \cdots & \cdots & \cdots & \boldsymbol{0}_m & \boldsymbol{B}^{*\prime}  & \boldsymbol{C}^{*}\\
\end{bmatrix},
\end{equation}
where, $\boldsymbol{A}^{*} = \boldsymbol{\Phi}^{*\prime} \boldsymbol{W}^{*+} \boldsymbol{\Phi}^{*}$, $\boldsymbol{B}^{*} = - \boldsymbol{\Phi}^{*\prime} \boldsymbol{W}^{*+}$ and $\boldsymbol{C}^{*} = \boldsymbol{W}^{*+}$ are $pm \times pm$ matrices and $\boldsymbol{W}^{*+}$ denotes the Moore-Penrose inverse.
Let $\boldsymbol{\Psi} = (\text{vec}(\boldsymbol{\Phi}_{1})^{\prime},\dots,\text{vec}(\boldsymbol{\Phi}_{p})^{\prime},\text{vec}(\boldsymbol{W})^{\prime})$ 
be the vector of parameters and let $\boldsymbol{x}= (\boldsymbol{x}_{1}^{*^\prime},\dots,\boldsymbol{x}_{T}^{*^\prime})^{\prime}$ be the observations from a VAR$(p)$ process. The joint p.d.f. is given by 
\begin{equation}\label{Eq: Joint pdf VAR_p}
    f(\boldsymbol{x}|\boldsymbol{\Psi}) = (2\pi)^{-pmT/2} |\boldsymbol{\Sigma}^{*}|^{-1/2} \exp\bigg({-\frac{1}{2} \boldsymbol{x}^{\prime}\boldsymbol{\Sigma}^{*-1}\boldsymbol{x}}\bigg).
\end{equation}

\subsubsection{Joint p.d.f. for a VARMA($p,q$) Process}


In the literature, we rarely see applications where the state evolution  in a dynamic model follows a VARMA process. Nevertheless,
for completeness, we derive the precision matrix for the joint p.d.f. of an $m$-dimensional time series following a stationary VARMA($p,q$) process
defined by
\begin{equation}\label{Eq: VARMA}
    \boldsymbol{\Phi}(B) \boldsymbol{x}_t = \boldsymbol{\Theta}(B) \boldsymbol{w}_{t},  
\end{equation}
where $\boldsymbol{\Phi}(B)$ is the VAR operator as defined in (\ref{Eq: VAR_operator}) and the moving average (MA) operator $\boldsymbol{\Theta}(B)$ is defined as 
\begin{equation*}\label{Eq: MA_operator}
    \boldsymbol{\Theta}(B) = \boldsymbol{I}_{m} - \boldsymbol{\Theta}_{1}B - \dots - \boldsymbol{\Theta}_{q}B^{q}.
\end{equation*}
In \eqref{Eq: VARMA}, $\boldsymbol{\Phi}_{i}$ and $\boldsymbol{\Theta}_{i}$ are $m \times m$ unknown parameter matrices
and the $m$-dimensional white noise process is represented by $\boldsymbol{w}_{t} = (w_{t,1},\dots,w_{t,m})$ with $E(\boldsymbol{w}_{t}) = \boldsymbol{0}$ and $Cov(\boldsymbol{w}_{t}) = \boldsymbol{\Sigma}_{w}$, which a p.d. variance-covariance matrix. Let $\boldsymbol{\Psi}^{*} = (\text{vec}(\boldsymbol{\Phi}_1)^{\prime},\dots,\text{vec}(\boldsymbol{\Phi}_{p})^{\prime},\text{vec}(\boldsymbol{\Theta}_1)^{\prime},\dots,\text{vec}(\boldsymbol{\Theta}_{q})^{\prime},\text{vec}(\boldsymbol{\Sigma}_w)^{\prime})^{\prime}$ be the vector of parameters and assuming Gaussian errors, the joint p.d.f. is given by


\begin{equation}
    L(\boldsymbol{\Psi}^{*}|\boldsymbol{x}) = (2\pi)^{-mT/2} |\boldsymbol{\Sigma}|^{-1/2} \exp\bigg(-\frac{1}{2} \boldsymbol{x}^{\prime}\boldsymbol{\Sigma}^{-1}\boldsymbol{x}\bigg),
\end{equation}
where $T$ is the number of observations, $\boldsymbol{x} = (\boldsymbol{x}^{\prime}_1,\dots,\boldsymbol{x}^{\prime}_{T})^{\prime}$ is $mT$-dimensional vector and $\boldsymbol{\Sigma}$ is the $mT \times mT$ variance-covariance matrix of the observations. Using the closed form expressions in \cite{gallego2009exact}, we have
\begin{equation}
    L(\boldsymbol{\Psi}^{*}|\boldsymbol{x}) = L(\boldsymbol{\Psi}^{*}|\hat{\boldsymbol{w}}_{0}) = (2\pi)^{-mT/2} |\boldsymbol{\Sigma}_0|^{-1/2} \exp \bigg(-\frac{1}{2} \hat{\boldsymbol{w}}^{\prime}_0 \boldsymbol{\Sigma}^{-1}_0 \hat{\boldsymbol{w}}_0\bigg),
\end{equation}
where $\boldsymbol{\Sigma}_0$ is the variance-covariance matrix of $\hat{\boldsymbol{w}}_0$. Let $\boldsymbol{u}=\left(\boldsymbol{x}_{-p+1}, \ldots, \boldsymbol{x}_{0}, \boldsymbol{w}_{-q+1}, \ldots, \boldsymbol{w}_{0}\right)$ is a $m(p+q) \times 1$ vector of pre-sample values and covariance of pre-sample values be given by $\boldsymbol{\Sigma}_u$. The block matrix $\boldsymbol{\Phi}=\left[\boldsymbol{\Phi}_{i j}\right](i=1, \ldots, T ; j=1, \ldots, T)$ is defined as $\boldsymbol{\Phi}_{i j}=-\boldsymbol{\Phi}_{i-j}$ for $0 \leq i-j \leq p$ and $\boldsymbol{\Phi}_{i j}=\mathbf{0}$ otherwise, where the $m \times m$ matrix $\boldsymbol{\Phi}_{i j}$ is the typical block of $\boldsymbol{\Phi}$ and $\boldsymbol{\Phi}_{0}=-\mathbf{I}_{m}$. Similarly, the block matrix $\boldsymbol{\Theta}=\left[\boldsymbol{\Theta}_{i j}\right](i=1, \ldots, T ; j=1, \ldots, T)$ is such that $\boldsymbol{\Theta}_{i j}=-\boldsymbol{\Theta}_{i-j}$ for $0 \leq i-j \leq q$ and $\boldsymbol{\Theta}_{i j}=\mathbf{0}$ otherwise, where $\boldsymbol{\Theta}_{0}=-\mathbf{I}_{m}$. Let $r=\max (p, q)$ and the block matrix $\mathbf{F}=\left[\mathbf{F}_{i j}\right](i=1, \ldots, r ; j=1, \ldots, p+q)$ is such that $\mathbf{F}_{i j}=\boldsymbol{\Phi}_{p-(j-i)}$ for $0 \leq j-i<p, \mathbf{F}_{i j}=-\boldsymbol{\Theta}_{p+q-(j-i)}$ for $p \leq j-i<p+q$, and $\mathbf{F}_{i j}=\mathbf{0}$ otherwise. The $k T \times k r$ matrix $\mathbf{G}$ has 1s along the main diagonal and 0s otherwise. Hence the precision matrix as in \cite{gallego2009exact} is given by
\begin{equation}
    \boldsymbol{\Sigma}_{0}^{-1}=\left(\mathbf{I}_{T} \otimes \boldsymbol{\Sigma}_{w}^{-1}\right)-\left(\mathbf{I}_{T} \otimes \boldsymbol{\Sigma}_{w}^{-1}\right) \boldsymbol{\Theta}^{-1} \mathbf{G F}\left(\mathbf{Z}^{\prime} \mathbf{Z}\right)^{-1} \mathbf{F}^{\prime} \mathbf{G}^{\prime} \Theta^{\prime-1}\left(\mathbf{I}_{T} \otimes \boldsymbol{\Sigma}_{w}^{-1}\right),
\end{equation}
where 
\begin{equation*}
    \mathbf{Z}^{\prime} \mathbf{Z}=\boldsymbol{\Sigma}_{u}^{-1}+\mathbf{F}^{\prime} \mathbf{G}^{\prime} \mathbf{\Theta}^{\prime-1}\left(\mathbf{I}_{T} \otimes \boldsymbol{\Sigma}_{w}^{-1}\right) \boldsymbol{\Theta}^{-1} \mathbf{G F}
\end{equation*}
is a square matrix of order $m(p+q)$.

\subsection{Approximate Posterior Analysis}\label{Prior}

Let $\boldsymbol{Y}$ and $\boldsymbol{S}$ respectively denote the responses $\boldsymbol{y}_{it}$ and the static predictors $\boldsymbol{S}^\prime_{j,it}$ for $i = 1,\dots,n$, $j=1,\dots,m$ and $t=1,\dots,T$. Let $\boldsymbol{b} = (\beta_{ji0},\boldsymbol{\beta}_{j,i}^{\prime})^{\prime}$ denote all the subject-specific intercepts $\beta_{ji0}$ and the fixed effects $\boldsymbol{\beta}_{j,i}$, for $i = 1,\dots,n$ and $j = 1,\dots,m$,  $\boldsymbol{\Phi} = (\text{vec}^{\prime}(\boldsymbol{\Phi}_{1}),\dots,\text{vec}^{\prime}(\boldsymbol{\Phi}_{p}))^{\prime}$ denote all the parameters related to the coefficient matrices in the latent states and let $\boldsymbol{x} = (\boldsymbol{x}^{\prime}_{1},\dots,\boldsymbol{x}^{\prime}_{T})^{\prime}$ denotes the latent states. Let $\tau$ be the precision parameter of the observation equation and $\boldsymbol{\theta}_{it} = (\theta_{1,it},\dots,\theta_{m,it})^{\prime}$, where $\theta_{j,it}$ evolves as in (\ref{Eq: Gamma Obs Eqn}) for $j=1,\dots,m$. The likelihood function under the model described by equations (\ref{Eq: Gamma marginal}) - (\ref{Eq: VAR_state}) is:
\begin{equation}\label{Eq: Likelihood LCM}
\begin{aligned}
    & L(\boldsymbol{b}, \boldsymbol{\Phi},\boldsymbol{W}, \boldsymbol{\Sigma},\boldsymbol{x},\tau|\boldsymbol{Y},\boldsymbol{S}) = \int \prod_{i = 1}^{n} \prod_{t = 1}^{T} p(\boldsymbol{y}_{it}|\boldsymbol{\theta}_{it},\tau) \times p(\boldsymbol{x}_{t}|\boldsymbol{x}_{t-1},\dots,\boldsymbol{x}_{t-p},\boldsymbol{W}) \times p(\boldsymbol{\alpha}_{it}|\boldsymbol{\Sigma}) \, d\boldsymbol{\alpha}_{it},
\end{aligned}
\end{equation}
where     
\begin{equation*}
    p(\boldsymbol{y}_{it}|\boldsymbol{\theta}_{it},\tau) = \prod_{j=1}^{m}\bigg( \dfrac{\tau^{\tau}}{\Gamma(\tau) \theta_{j,it}^{\tau}} y_{j,it}^{\tau -1} \exp\Big(-\frac{\tau}{\theta_{j,it}}y_{j,it}\Big)\bigg),
\end{equation*}
\begin{equation*}
    p(\boldsymbol{\alpha}_{it}|\boldsymbol{\Sigma}) = \dfrac{1}{\sqrt{(2\pi)^{m} \det(\boldsymbol{\Sigma})}} \exp \bigg(-\dfrac{1}{2} \Big(\boldsymbol{\alpha}_{it}^\prime \boldsymbol{\Sigma}^{-1}\boldsymbol{\alpha}_{it}\Big)\bigg)
\end{equation*}
and 
$\prod_{t=1}^{T}p(\boldsymbol{x}_{t}|\boldsymbol{x}_{t-1},\dots,\boldsymbol{x}_{t-p},\boldsymbol{W})$ is the joint p.d.f. of VAR$(p)$ process and hence can be computed as in (\ref{Eq: Joint pdf VAR_p}).



The LCM model parameters
include the subject-specific intercepts $\beta_{j,i0}$, the fixed effects $\boldsymbol{\beta}_{j,i}$ -- a $p_{j}$ dimensional vector corresponding to $\boldsymbol{S}^{\prime}_{j,it}$, the $m \times m$ variance-covariance matrix of the random effects $\boldsymbol{\Sigma}$ and the $m \times m$ variance-covariance matrix of the state errors $\boldsymbol{W}$. We have mostly used the default prior specifications of \texttt{R-INLA} which seem to work well. A Wishart distribution is considered as a prior specification for the $m \times m$ covariance matrix $\boldsymbol{\Sigma}$. In particular, the precision matrix $\boldsymbol{\Sigma}^{-1} \sim \text{Wishart}_{m}(r,\boldsymbol{I}_m)$, $r$ is the degree of freedom, whose default value is $2m+1$ representing the default prior specification for $m$-variate Gaussian random effects in \texttt{R-INLA}. For modeling purposes we have used a diagonal matrix specification for $\boldsymbol{W}$ = $\text{diag}(\sigma^2_{w1},\dots,\sigma^2_{wm})$, where $\sigma^2_{wj}$ are marginal variances for $j=1,\dots,m$. Let the corresponding log-precision be given by $\theta_{wj} = \log\bigg(\dfrac{1}{\sigma^2_{wj}}\bigg)$. Then for each $j$, the logarithm of the precision $\theta_{wj}$ is assumed to follow a log-gamma distribution, whose shape and inverse scale parameters are 1 and 0.00005 respectively. This is also the default prior specification for \texttt{R-INLA}. The prior for the fixed effects is chosen to be the default Gaussian prior with mean 0 and precision 0.001.

Given the data and model parameters, the Bayesian formulation requires specification of the likelihood and prior, from which, using Bayes' Theorem, we obtain the posterior density as a normalized product of the likelihood and prior. Since the likelihood (\ref{Eq: Likelihood LCM}) is not available in closed form, its exact analysis is cumbersome. 
The high computational requirements of a fully Bayesian approach has prompted us to use \texttt{R-INLA} for model fitting \citep{rue2009approximate,lindgren2015bayesian}. For approximate Bayesian inference, INLA uses nested Laplace approximations and numerical integration, thereby significantly reducing the computational time compared to a fully Bayesian inference via MCMC. 


One of the main advantages of INLA is that templates for many useful models in the literature are available in the  \texttt{R-INLA} package.  Other cases can be programmed and implemented using generic functions. We provide more details about the implementation of VAR($1$) as a latent process using \texttt{rgeneric} in \texttt{R-INLA} in the Appendix (\ref{appendix}). In the following section we show accurate estimation of parameters using simulated data from trivariate LCM-AR and trivariate LCM-VAR and then we fit trivariate LCMs to model interdependencies between the realized volatility measures for several stock indexes.


\section{Simulated Data}\label{Simulated Study}
Data from the trivariate LCM-AR model with $Y_{j,it}$ ($i=1,\dots,30$; $j=1,2,3$ and $t=1,\dots,500$) is simulated according to the following model:
\begin{equation}\label{Eq: LCM-AR_sim}
\begin{aligned}
    &Y_{j,it} | \theta_{j,it}, \tau \sim \text{Gamma}\bigg(\tau, \frac{\tau}{\theta_{j,it}}\bigg)\\
    &\log(\theta_{j,it}) = x_{j,t} + \alpha_{j,it} + \beta S_{j,it},
\end{aligned}    
\end{equation}
where $S_{j,it}$ represents a static predictor which is simulated from $N(0,1)$ and $\beta = 0.2$. The state equations are given by (\ref{AR state eq}) with the order of autoregression $p = 1$, $\phi_{1,1} = \phi_{2,1} = \phi_{3,1} = 0.8$ and the state errors $w_{1,t}, w_{2,t}$ and $w_{3,t}$ are simulated from $N(0,W_{1}), N(0,W_{2}) \text{ and } N(0,W_{3})$ respectively, where $W_{1} = 0.3$, $W_{2} = 0.2$ and $W_{3} = 0.5$. The level correlated term $\boldsymbol{\alpha}_{it}$ is simulated from $N_3(\boldsymbol{0},\boldsymbol{\Sigma})$, where 
\begin{equation}\label{Eq: Sigma}
\boldsymbol{\Sigma}=\begin{pmatrix}
\sigma^2_1 & \rho_{12} \sigma_1 \sigma_2 & \rho_{13} \sigma_1 \sigma_3 \\
\rho_{12} \sigma_1 \sigma_2 & \sigma^2_2 & \rho_{23} \sigma_2 \sigma_3\\
\rho_{13} \sigma_1 \sigma_3 & \rho_{23} \sigma_2 \sigma_3 & \sigma^2_3\\
\end{pmatrix},
\end{equation}
is the variance-covariance matrix with $\sigma^2_{1} = \sigma^2_{2}=\sigma^2_{3} = 0.5$ and $\rho_{12} = 0.8$, $\rho_{13} = 0.7$ and $\rho_{23} = 0.5$. The responses are simulated from gamma distribution with $\tau=300$. For the model estimation, we used the \texttt{R-INLA} package \cite{rue2009approximate,martins2013bayesian} with the default prior specification for the parameters as discussed in subsection \ref{Prior}. We assume a  normal prior for $\beta$, trivariate inverse Wishart prior for the variance-covariance matrix $\boldsymbol{\Sigma}$ and gamma prior for the precisions. We explicitly assumed $\tau \sim Gamma(0.01,0.01)$ i.e. E[$\tau$] = 1 and Var[$\tau$]=100. 

Simulations were conducted on a personal laptop (11th Gen Intel(R) Core(TM) i7-1165G7 @ 2.80GHz) with 16Gb of RAM using  64-bit version of Windows 11 operating system. It took about 7 minutes to run LCM-AR on the simulated data. Most of the estimated parameters in LCM-AR are close to their true values, except for $\tau$, the precision parameter of the observational equation, whose estimate is a bit off. Table \ref{LCM_AR simulation} gives the posterior summaries for the estimated parameters of LCM-AR. 
\begin{table}[H]
\caption{Posterior summaries for LCM-AR simulated data}
\fontsize{8}{10}
\label{LCM_AR simulation}
\begin{center}
\begin{tabular}{|c|c|c|c|}
\hline
Parameter  & True value & Posterior Mean & Posterior Std. Dev. \\ \hline
$\tau$  & 300   & 455.422 & 109.925 \\
$\beta$ & 0.200 & 0.206   & 0.003    \\ 
$1/\sigma^2_{1}$ & 2.000  & 1.962  & 0.046 \\
$1/\sigma^2_{2}$ & 2.000  & 1.909  & 0.037 \\
$1/\sigma^2_{3}$ & 2.000  & 1.875  & 0.049 \\
$\rho_{12}$ & 0.800  & 0.795 & 0.009 \\
$\rho_{13}$ & 0.700  & 0.728 & 0.011 \\
$\rho_{23}$ & 0.500  & 0.515 & 0.011 \\
$\phi_{11}$ & 0.800  & 0.730 & 0.010 \\
$\phi_{22}$  & 0.800 & 0.819 & 0.008 \\
$\phi_{33}$  & 0.800 & 0.807 & 0.007 \\
$1/\sigma^2_{w1}$  & 3.333 & 3.312	& 0.044 \\
$1/\sigma^2_{w2}$  & 5.000	&5.438	&0.059\\ 
$1/\sigma^2_{w3}$  & 2.000	& 2.149	& 0.021\\ \hline
\end{tabular}
\end{center}
\end{table}

Data from the trivariate LCM-VAR model with $Y_{j,it}$ ($i=1,\dots,30$; $j=1,2,3$ and $t=1,\dots,500$) is simulated according to observation equation as in (\ref{Eq: LCM-AR_sim}) and the state equation is described by 
\begin{equation*}
    \boldsymbol{x}_t = \boldsymbol{\Phi} \boldsymbol{x}_{t-1} + \boldsymbol{w}_t, 
\end{equation*}
where \begin{equation*}
\boldsymbol{\Phi}=\begin{pmatrix}
0.5 & 0 & 0.3 \\
0.6 & 0.1 & 0.5 \\
0.1 & 0 & 0.8 \\
\end{pmatrix}
\end{equation*}
and $\boldsymbol{w}_t$ is simulated from $N_{3}(\boldsymbol{0},\boldsymbol{W})$, where $\boldsymbol{W}= \text{diag}(0.5,0.25,0.5)$, i.e., $W_{1} = 0.5$, $W_{2} = 0.25$ and $W_{3} = 0.5$. The static predictor is simulated from $N(0,1)$ and the coefficient $\beta = 0.2$. The level correlated term $\boldsymbol{\alpha}_{it}$ is simulated from $N_3(\boldsymbol{0},\boldsymbol{\Sigma})$, where $\boldsymbol{\Sigma}$ is defined as in (\ref{Eq: Sigma})
with $\sigma^2_{1} = 1/3$, $\sigma^2_{2}= 1/2$,$\sigma^2_{3} = 1/3$ and $\rho_{12} = 0.6$, $\rho_{13} = 0.8$ and $\rho_{23} = 0.8$. The responses are simulated from gamma distribution with $\tau=100$. 

It took about 38 minutes to run LCM-VAR on the simulated data. Table \ref{LCM-VAR simulation} gives the posterior summaries for the estimated parameters of LCM-VAR. Similar to the LCM-AR simulation setup, most of the estimated parameters are close to their respective true values. The posterior standard deviation of the precisions of the state errors in LCM-VAR seems to be more than that in LCM-AR.

\begin{table}[H]
\caption{Posterior summaries for LCM-VAR simulated data}
\fontsize{8}{10}
\begin{center}
\label{LCM-VAR simulation}
\begin{tabular}{|c|c|c|c|}
\hline
Parameter & True value & Posterior Mean & Posterior Std. Dev. \\ \hline
$\tau$               & 100               & 83.321                   & 10.360                          \\ 
$\beta$               & 0.200               & 0.201                   & 0.002                          \\ 
$1/\sigma_1^2$            & 3.000               & 3.109                   & 0.038                          \\ 
$1/\sigma_2^2$            & 2.000               & 2.025                   & 0.023                          \\ 
$1/\sigma_3^2$            & 3.000               & 3.115                   & 0.038                          \\ 
$\rho_{12}$          & 0.600               & 0.644                   & 0.007                          \\ 
$\rho_{13}$          & 0.800               & 0.847                   & 0.006                          \\ 
$\rho_{23}$          & 0.800               & 0.843                   & 0.005                          \\ 
$\phi_{11}$          & 0.500               & 0.481                   & 0.049                          \\ 
$\phi_{21}$          & 0.600               & 0.569                   & 0.030                          \\ 
$\phi_{31}$          & 0.100               & 0.013                   & 0.036                          \\ 
$\phi_{12}$          & 0.000               & -0.006                  & 0.042                          \\ 
$\phi_{22}$          & 0.100               & 0.090                   & 0.031                          \\ 
$\phi_{32}$          & 0.000               & -0.041                  & 0.039                          \\ 
$\phi_{13}$          & 0.300               & 0.289                   & 0.049                          \\ 
$\phi_{23}$          & 0.500               & 0.517                   & 0.025                          \\ 
$\phi_{33}$          & 0.800               & 0.855                   & 0.028                          \\ 
$1/\sigma^2_{w1}$            & 2.000               & 1.783                   & 0.095                          \\ 
$1/\sigma^2_{w2}$            & 4.000               & 3.557                   & 0.207                          \\ 
$1/\sigma^2_{w3}$            & 2.000               & 2.314                   & 0.124                          \\ \hline
\end{tabular}
\end{center}
\end{table}

\section{Modeling Multivariate Realized Volatility Measures}\label{Empirical Results}

Volatility measurement is central to asset pricing, asset allocation, portfolio optimization and risk management. A first attempt to measure volatility using intraday data was made by \cite{parkinson1980extreme} for the estimation of the daily
range. Since then the literature has expanded significantly from the realized volatility of \cite{andersen1998answering} to bipower variation \citep{barndorff2004power}, realized kernels of \cite{barndorff2009realized} and realized covariance matrices \citep{ait2010high,peluso2014bayesian}. As these intraday volatility measures evolved, there has been a natural complementary effort to build adequate models to describe their joint dynamics \citep{engle2006multiple,donelli2021bayesian,cipollini2013semiparametric}.

\subsection{Data description}
The focus of this study is to model three robust intraday volatility measures: median realized variance ($MedRV$), realized kernel variance based on Parzen kernel ($RK$) and bipower variation ($BPV$) using the different level correlated model specifications for the $30$ stock indexes data (as shown in Table~\ref{tab:data}) available on Oxford-Man Institute's Quantitative Finance Realized Library \citep{heber2009oxford}. These volatility measures are based on $5$-minute sampling frequency.
The sample period spans from January $2$, $2015$, to December $29$, $2017$.  The number of observations for each stock index is 513. Thus we have $m = 3$, $n=30$ and $T=513$. In Table (\ref{tab:data}), we list all the stock indexes we have used in our analysis along with their associated regions.
\begin{table}[H] 
\begin{center}
\caption{List of stock indexes along with their associated regions}
\footnotesize
\begin{tabular}{|l|l|l|}
\hline
\textbf{Symbol} & \textbf{Name}                             & \textbf{Region} \\ \hline
AEX            & AEX index                                 & Netherlands      \\ 
AORD           & All Ordinaries                            & Australia        \\ 
BFX            & Bell 20 Index                             & Belgium          \\ 
BSESN          & S\&P BSE Sensex                           & India            \\ 
BVSP           & BVSP BOVESPA Index                        & United States    \\ 
DJI            & Dow Jones Industrial Average              & United States    \\ 
FCHI           & CAC 40                                    & France           \\ 
FTMIB          & FTSE MIB                                  & Italy            \\ 
FTSE           & FTSE 100                                  & United Kingdom   \\ 
GDAXI          & DAX                                       & Germany          \\ 
GSPTSE         & S\&P/TSX Composite index                  & Canada           \\ 
HSI            & HANG SENG Index                           & Hong Kong        \\ 
IBEX           & IBEX 35 Index                             & United States    \\ 
IXIC           & Nasdaq 100                                & United States    \\ 
KS11           & Korea Composite Stock Price Index         & South Korea      \\ 
KSE            & Karachi SE 100 Index                      & Pakistan         \\ 
MXX            & IPC Mexico                                & Mexico           \\ 
N225           & Nikkei 225                                & Japan            \\ 
NSEI           & NIFTY 50                                  & India            \\ 
OMXC20         & OMX Copenhagen 20 Index                   & Denmark          \\ 
OMXHPI         & OMX Helsinki All Share Index              & Finland          \\ 
OMXSPI         & OMX Stockholm All Share Index             & Sweden           \\ 
OSEAX          & Oslo Exchange All-share Index             & Norway           \\ 
RUT            & Russel 2000                               & United States    \\ 
SMSI           & Madrid General Index                      & Spain            \\ 
SPX            & S\&P 500 Index                            & United States    \\ 
SSEC           & Shanghai Composite Index                  & China            \\ 
SSMI           & Swiss Stock Market Index                  & Switzerland      \\ 
STI            & Straits Times Index                       & Singapore        \\ 
STOXX50E       & EURO STOXX 50                             & Europe           \\ \hline
\end{tabular}
\label{tab:data}
\end{center}
\end{table} 

The mean, standard deviation (St.Dev.), skewness, kurtosis and autocorrelation of lag 1 of $\sqrt{MedRV}$, $\sqrt{RK}$ and $\sqrt{BPV}$ are given in Table (\ref{tab:summary}). The values of skewness indicates that all of the realized measures are positively skewed and high values of kurtosis indicate the presence of heavy tails. All of the realized measures are persistent according to the high values of their first order autocorrelation.

\begin{table}[H]
\begin{center}
\captionsetup{width=.7\textwidth}
\caption{Descriptive Statistics of $\sqrt{MedRV}$, $\sqrt{RK}$ and $\sqrt{BPV}$}
\begin{tabular}{|c|c|c|c|c|c|}
\hline
& Mean  & St.Dev. & Skewness & Kurtosis & AutoCorr(lag1) \\ \hline
$\sqrt{MedRV}$      & 0.004 & 0.003  & 2.314    & 13.094   & 0.750           \\ $\sqrt{RK}$ & 0.007 & 0.005  & 3.487    & 30.008   & 0.583          \\ 
$\sqrt{BPV}$         & 0.007 & 0.004  & 3.699    & 32.181   & 0.706  \\
\hline
\end{tabular}
\label{tab:summary}
\end{center}
\end{table}

We compute the kendall's correlation matrix for the components of the response vector in Table \ref{tab: kendall's tau}. The off-diagonal elements indicate high degree of positive correlation between the realized volatility measures.

\begin{table}[H]
\begin{center}
\captionsetup{width=.7\textwidth}
\caption{Kendall's correlation matrix of $\sqrt{MedRV}$, $\sqrt{RK}$ and $\sqrt{BPV}$}
\label{tab: kendall's tau}
\begin{tabular}{|c|c|c|c|}
\hline
 & $\sqrt{MedRV}$ & $\sqrt{RK}$ & $\sqrt{BPV}$ \\ \hline
$\sqrt{MedRV}$       & 1.00          & 0.62             & 0.77     \\
$\sqrt{RK}$    & 0.62          & 1.00             & 0.72     \\ 
$\sqrt{BPV}$            & 0.77          & 0.72             & 1.00     \\ \hline
\end{tabular}
\end{center}
\end{table}

\subsection{Realized Volatility Measures} \label{Data Description}
Let $\{P_{t_{j}}\}$ be a sequence of observed intraday prices of an asset. Consider a time interval $[a,b]$ and the time index $t_j$ for $t_j \in [a,b]$. The time is partitioned into $M$ equidistant points in time $j=1,\dots,M$. At each time point $t_j$, the price of an asset is observed. Then $\{P_{t_{j}}\}_{j=1}^{M}$ is the observed sequence of prices of an asset sampled at frequency $\tau=\frac{b-a}{M-1}$. Hence the intraday returns over the sub-interval $[t_{j-1},t_j)$ is defined as 
$r_{j} = \log(P_{t_j}) - \log(P_{t_{j-1}})$. 
To obtain realized variances at 5-minutes sampling frequency, $\tau = 300$ seconds and similarly for 1-minute sampling frequency, $\tau = 60$ seconds. 

Let us denote $r_{j,t}$ as the $j^{th}$ intraday return for $t^{th}$ day, for $t=1,\dots,T$. 
We define some popular measures of intraday volatility considered in the literature.
Median realized variance ($MedRV$) is an estimator introduced by \cite{andersen2012jump}, which is a scaled square of the median of three consecutive intraday absolute returns, which is calculated as
\begin{equation*}
    \text{MedRV}_t = \dfrac{\pi}{6-4\sqrt{3}+\pi} \dfrac{M}{M-2}\sum_{j=2}^{M-1} \text{median}\big(|r_{j-1,t}|,|r_{j,t}|,|r_{j+1,t}|\big)^2.
\end{equation*}
\cite{barndorff2009realized} proposed the realized kernels estimator, which uses kernel methods to combat market microstructure noise. The realized kernel estimator is
\begin{equation*}
\begin{aligned}
    \text{RK}_t = \sum_{h= -H}^{H} k\Big(\frac{h}{H+1}\Big) \gamma_h, \quad \text{where}\\
    \gamma_h = \sum_{i=|h|+1}^{n} r_{i,t} r_{i-|h|,t},
\end{aligned}
\end{equation*}
where $k(.)$ is the Parzen kernel function defined as 
\[ k(x)= \begin{cases} 
      1-6x^2 + 6x^3 & 0\leq x\leq 1/2 \\
      2(1-x)^3 & 1/2\leq x\leq 1 \\
      0 & x>1 .
   \end{cases}
\]
A simple and comprehensible measure, which isolates the variation of the pure price process is the Bipower variation (BPV). It is calculated as the sum of products of absolute values of two consecutive returns,
\begin{equation*}
    \text{BPV}_t = \sum_{j=2}^{M} |r_{j,t}| |r_{j-1,t}|.
\end{equation*} 
For each day $t$, we consider the $m=3$-dimensional time series composed of standard deviation versions of median realized variance, realized kernel and bipower variation, $\boldsymbol{y}_t =(\sqrt{RV}_t,\sqrt{RK}_t,\sqrt{BPV}_t )^\prime$, where $t= 1,\dots,T.$ We have plotted the time series of $\sqrt{MedRV}$, $\sqrt{RK}$ and $\sqrt{BPV}$ for two stock indexes S\&P 500 and NASDAQ-100 for illustration.

\begin{figure}[H]
    \centering
    \caption{Plots of $\sqrt{MedRV}$, $\sqrt{RK}$ and $\sqrt{BPV}$ for SP500 and NASDAQ-100 from January 2, 2015, to December 29, 2017}
    \includegraphics[scale=0.5]{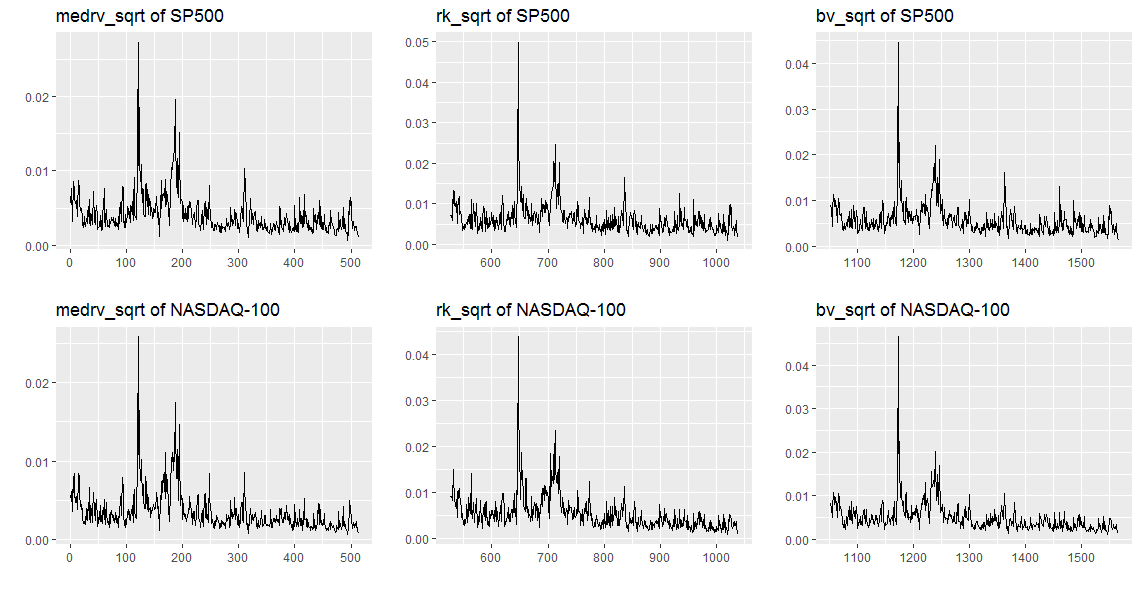}
    \label{fig:my_label}
\end{figure}


\subsection{Model Specification}\label{Models}

Let $\boldsymbol{y}_{it} = (y_{1,it},\dots,y_{m,it})^\prime$ be a $m$-variate vector of positive-valued volatility measures over time, for $i = 1 ,\ldots,n$ and $t = 1 ,\ldots,T$. That is, we observe $m$ components of different volatility measures on $n$ stock market indexes over $T$ regularly spaced time points. 
High values of the lag 1 autocorrelation has prompted us to use 
$\log (Y_{j,i,t-1})$ as the predictor indicating the daily level component driven by short term traders as in \cite{corsi2009simple}. We define the daily jump component as $J_t = \max(RV_t - BPV_t,0)$ and the daily continuous component as $C_t = RV_t - J_t$, where $RV_t = \sum_{j=1}^{M} r_{j,t}^2$, is the realized variance at time $t$. The daily jump and continuous components account for the jumps in the price process,  and are useful predictors in modeling realized volatility \citep{andersen2007roughing}.
With these predictors,  (\ref{Eq: Gamma marginal}) -- (\ref{Eq: Gamma Obs Eqn}) simplify as
\begin{equation} \label{Eq: LCM1}
\begin{aligned}
 &Y_{j,it} | \theta_{j,it}, \tau \sim   \text{Gamma}\bigg(\tau, \frac{\tau}{\theta_{j,it}}\bigg) ,\\
 \log(\theta_{j,it}) = \alpha_{j,i} + \beta_{j,t,0} + & \beta_{1,j} \log (Y_{j,i,t-1}) + \beta_{2,j} \log(1+ J_{j,i,t-1})+ \beta_{3,j} \log(1+ C_{j,i,t-1}) + \xi_{j,it}, \\
\end{aligned}
\end{equation}
where the level $\alpha_{j,i}$ is assumed to be stock-specific, and 
the level correlated effect $\boldsymbol{\xi} = (\xi_{1,it},\xi_{2,it},\xi_{3,it})^{\prime} \sim N(\boldsymbol{0},\boldsymbol{\Sigma})$, with 
\begin{equation*}\label{Eq: correlation}
\boldsymbol{\Sigma}=\begin{pmatrix}
\sigma^2_1 & \rho_{12} \sigma_1 \sigma_2 & \rho_{13} \sigma_1 \sigma_3 \\
\rho_{12} \sigma_1 \sigma_2 & \sigma^2_2 & \rho_{23} \sigma_2 \sigma_3\\
\rho_{13} \sigma_1 \sigma_3 & \rho_{23} \sigma_2 \sigma_3 & \sigma^2_3\\
\end{pmatrix}.
\end{equation*}


We refer to this model as LCM1, and
consider two different specifications for the time-varying intercept $\beta_{j,t,0}$,
similar to (\ref{Eq: VAR_state}) and (\ref{AR state eq}). 
First, an assumption of VAR evolution leads to 
\begin{equation}\label{Eq: VAR_evolution}
    \boldsymbol{\beta}_{t,0} = \boldsymbol{\Phi} \boldsymbol{\beta}_{t-1,0} + \boldsymbol{w}_t,  \text{ where }
    \boldsymbol{w}_t \sim N(\boldsymbol{0},\boldsymbol{W}),
\end{equation}
while an AR specification for each component $j = 1, \ldots, m$ leads to
\begin{equation}\label{Eq: AR_evolution}
    \beta_{j,t,0} = \phi_{j}\beta_{j,t-1,0} + w_{j,t}, \text{ where } w_{j,t} \sim N(0, W_{j}).
\end{equation}
We refer to  (\ref{Eq: LCM1}) and (\ref{Eq: VAR_evolution}) as LCM1-VAR, and to (\ref{Eq: LCM1}) and (\ref{Eq: AR_evolution}) as LCM1-AR.

We also investigate a setup where the level $\alpha_{j,i}$ is component-specific but \textit{not} stock-specific, and define 
the corresponding model LCM2 as
\begin{equation} \label{Eq: LCM2}
\begin{aligned}
 Y_{j,it} | \theta_{j,it} \sim  &  \quad \text{Gamma}\bigg(\tau, \frac{\tau}{\theta_{j,it}}\bigg) ,\\
  \log(\theta_{j,it}) = \alpha_{j} + \beta_{j,t,0} + & \beta_{1,j} \log (Y_{j,i,t-1}) + \beta_{2,j} \log(1+ J_{j,i,t-1})+ \beta_{3,j} \log(1+ C_{j,i,t-1}) + \xi_{j,it}, \\
\end{aligned}
\end{equation}
In this case, 
we refer to equations (\ref{Eq: LCM2}) and (\ref{Eq: VAR_evolution}) as LCM2-VAR, and to (\ref{Eq: LCM2}) and (\ref{Eq: AR_evolution}) as LCM2-AR. 
The difference between LCM1 and LCM2 lies in the specification of the level $\alpha$.

\subsection{Results}\label{Results}
Using \texttt{R-INLA}, we fit the four models discussed in Section (\ref{Models}): LCM1-VAR, LCM1-AR, LCM2-VAR, and LCM2-AR, using the same prior specification as mentioned in Section (\ref{Prior}). We have 
assumed $\tau \sim \text{Gamma}(1,0.1)$ such that $\text{E}[\tau] = 10$ and $\text{Var}[\tau]=100$. 



In Table \ref{Table:Level}, 
we show the mean of the posterior means of the levels from Model LCM1. We also show the posterior mean of the levels from Model LCM2. 

\begin{table}[H]
\small
\caption{Posterior summaries of component-specific levels}
\label{Table:Level}
\begin{center}
\begin{tabular}{|c| c| c| c| c|}
\hline
 & LCM1-VAR  & LCM1-AR & LCM2-VAR & LCM2-AR \\ \hline
Parameter & Mean of Posterior Mean & Mean of Posterior Mean & Posterior Mean & Posterior Mean \\ \hline
$\alpha_{1}$ & -4.401 & -4.898 & -3.312 & -3.869 \\ 
$\alpha_{2}$ & -4.443 & -4.633 & -3.657 & -3.923 \\ 
$\alpha_{3}$ & -3.864 & -4.450 & -3.062 & -3.583 \\ \hline
\end{tabular}
\end{center}
\end{table}

In Table \ref{Table: Covariates and Hyperparameter}, we show posterior summaries of the covariate coefficients and hyperparameters for Models LCM1 and LCM2. 
We observe that the $\beta$'s corresponding to the daily effects are higher under the AR specification than those obtained by assuming a VAR(1) state evolution, whereas the estimated state precisions are higher for the VAR(1) state evolution than the AR(1) state evolution. 
The estimated level correlations among the three volatility measures, i.e., $\rho_{12},\rho_{13}$ and $\rho_{23}$, across all the models are strong and significant, which empirically confirms dependency between them. The estimated level correlations from the four models are close to the Kendall's tau as discussed in Table (\ref{tab: kendall's tau}). Most of the estimated parameters are significant except for the jump components.



\begin{table}[H]
\small
\caption{Posterior summaries of covariate coefficients and hyperparameters}
\label{Table: Covariates and Hyperparameter}
\begin{center}
\begin{tabular}{|c|c|c|c|c|}
\hline
 & LCM1-VAR & LCM1-AR & \multicolumn{1}{l|}{LCM2-VAR} & \multicolumn{1}{l|}{LCM2-AR} \\ \hline
Parameter & Posterior Mean & Posterior Mean & Posterior Mean & Posterior Mean \\ \hline
\textbf{Covariate coefficients} & & & & \\ 
$\beta_{1,1}$ & 0.206 & 0.120 & 0.405 & 0.306 \\ 
$\beta_{1,2}$ & 0.122 & 0.088 & 0.281 & 0.229 \\ 
$\beta_{1,3}$ & 0.244 & 0.133 & 0.404 & 0.303 \\ 
$\beta_{2,1}$ & 1.661 & 42.829 & 5.444 & 36.690 \\ 
$\beta_{2,2}$ & 14.860 & 31.900 & -23.530 & 8.912 \\ 
$\beta_{2,3}$ & 21.604 & 57.976 & 6.873 & 37.788 \\ 
$\beta_{3,1}$ & 263.115 & 307.014 & 220.804 & 295.162 \\ 
$\beta_{3,2}$ & 227.402 & 273.007 & 249.234 & 291.896 \\ 
$\beta_{3,3}$ & 183.556 & 245.000 & 186.188 & 268.420 \\ 
\hline
\textbf{Hyperparameters} & & & & \\ 
$1/\sigma_{1}^2$ & 13.034 & 11.177 & 11.472 & 9.030 \\ 
$1/\sigma_{2}^2$ & 9.128 & 7.431 & 7.672 & 6.649 \\ 
$1/\sigma_{3}^2$ & 17.758 & 13.290 & 15.152 & 11.019 \\ 
$\rho_{12}$ & 0.621 & 0.677 & 0.605 & 0.676 \\ 
$\rho_{13}$ & 0.811 & 0.854 & 0.802 & 0.859 \\ 
$\rho_{23}$ & 0.807 & 0.837 & 0.798 & 0.841 \\ 
$\phi_{11}$ & 2.419 & 0.996 & 0.134 & 0.997 \\ 
$\phi_{21}$ & 2.564 &  & 0.309 &  \\ 
$\phi_{31}$ & 2.464 &  & 0.444 &  \\ 
$\phi_{12}$ & 0.592 &  & 2.482 &  \\ 
$\phi_{22}$ & 0.623 & 0.996 & 2.496 & 0.998 \\ 
$\phi_{32}$ & 0.526 &  & 2.362 &  \\ 
$\phi_{13}$ & -2.318 &  & -2.205 &  \\
$\phi_{23}$ & -2.382 &  & -2.246 &  \\ 
$\phi_{33}$ & -2.359 & 0.997 & -2.410 & 0.998 \\ 
$1/W_{1}$ & 428.255 & 20.633 & 768.665 & 32.634 \\ 
$1/W_{2}$ & 173.920 & 21.040 & 458.162 & 29.387 \\ 
$1/W_{3}$ & 667.143 & 25.496 & 868.763 & 41.081 \\ \hline
\end{tabular}
\end{center}
\end{table}

\subsection{Model Comparisons}
In volatility modeling, we are mostly interested in out-of-sample forecasting accuracy. We 
compare the four models discussed in section \ref{Results} using popular measures such as mean absolute percentage error (MAPE) and mean absolute error (MAE). Let $y_{n_{t}}(l)$ be the $l^{th}$ step-ahead forecast from origin $n_{t}$ for $y_{t+l}$, $l=1,\ldots,n_{h}$, and let $e_{n_{t}}(l) = y_{n_{t}+l} - y_{n_{t}}(l)$ be the $l^{th}$ step-ahead forecast error. MAPE is defined as
\begin{equation*}
    \text{MAPE} = \frac{1}{n_{h}} \sum_{l=1}^{n_h} |e_{n_{t}}(l)/y_{n_{t}+l}|.
\end{equation*}
MAE is defined as
\begin{equation*}
    \text{MAE} = \frac{1}{n_{h}}\sum_{l=1}^{n_{h}}|e_{n_{t}}(l)|.
\end{equation*}
Small values of MAPE and MAE indicate better model. To evaluate forecast accuracy, for each stock index and each realized volatility measure we calculated both MAPE and MAE based on 1-step, 5-step and 10-step ahead forecasts i.e. for $n_{h} = 1,5$ and $10$. In Table (\ref{Table: Forecast accuracy}), we compute MAPE and MAE as the average across stock indexes and across the realized volatility measures.

\begin{table}[H]
\begin{center}
\captionsetup{width=.7\textwidth}
\caption{MAPE and MAE averaged across the 30 stock indexes and also across 3 realized volatility measures}
\label{Table: Forecast accuracy}
\begin{tabular}{|c|c|c|c|c|}
\hline
Horizon ($n_h$) & LCM1-AR & LCM1-VAR & LCM2-AR & LCM2-VAR \\ \hline
\textbf{MAPE} & & & &\\
1 & 0.302 & 0.320 & 0.310 & $\boldsymbol{0.282}$ \\ 
5 & 0.307 & 0.322 & 0.318 & $\boldsymbol{0.252}$ \\ 
10 & 0.317 & 0.367 & 0.324 & $\boldsymbol{0.280}$ \\ \hline
\textbf{MAE} & & & &\\
1 & 0.0012 & 0.0013 & 0.0013 & $\boldsymbol{0.0011}$ \\ 
5 & 0.0012 & 0.0014 & 0.0012 & $\boldsymbol{0.0010}$ \\ 
10 & 0.0013 & 0.0016 & 0.0012 & $\boldsymbol{0.0011}$ \\ \hline
\end{tabular}
\end{center}
\end{table}
Thus, based on MAPE and MAE, LCM2-VAR outperforms all other models. For in-sample comparison we have used Deviance Information Criterion (DIC) \citep{spiegelhalter2002bayesian} as the model selection criterion. In that case, LCM1-AR was outperforming the rest due to smaller values of DIC, but this model does not perform well in terms of forecasting accuracy as in Table \ref{Table: Forecast accuracy}.

\section{Summary and discussion}\label{Discussion}

This article describes the use of the \texttt{rgeneric()} function in \texttt{R-INLA} for fitting an LCM with VAR$(p)$ latent effects 
to vector positive-valued time series. We demonstrate the method using simulated data and realized financial volatilities.
The proposed LCM-VAR framework is not limited to financial applications, or to temporal models with exogenous predictors. We can easily adapt the model to incorporate spatial dependence between different locations, with applications in modeling climate variables.

Another useful direction 
is a hierarchical multivariate dynamic model (HMDM) setup where the observed data vector $\boldsymbol{y}_{it} = (Y_{1,it},\dots,Y_{m,it})^\prime$ follows a multivariate Gamma distribution. 
Several multivariate extensions of the univariate gamma distributions exist in the literature (see \cite{cherian1941bi}, \cite{ramabhadran1951multivariate}, \cite{mathai1991multivariate}, and  \cite{kotz2004continuous} and references therein). 
Here we define multivariate gamma distribution as defined in \citep{tsionas2004bayesian} which is in the same spirit as in \citep{mathai1991multivariate}. 
Following \citep{mathai1991multivariate} and \citep{tsionas2004bayesian} to define the probability distribution function (p.d.f.) of an $m$-variate gamma distribution, we assume that 
\begin{equation*}\label{Eq: MG}
    Y_{j,it} = V_{j,it} + Z,
\end{equation*}
where $V_{j,it}$ and $Z$ are independent random variables distributed as Gamma$(\lambda_{j,it},\theta_{j,it})$, $i=1,\dots,n$, $j=1,\dots, m$, $t=1,\dots,T$ and Gamma$(\alpha,\gamma)$ respectively. 
Then for $i=1,\dots,n$, $j=1,\dots,m$ and $t=1,\dots, T$,
\begin{equation*}
    p(y_{j,it}|z,\lambda_{j,it},\theta_{j,it}) = \dfrac{\theta_{j,it}^{\lambda_{j,it}}}{\Gamma(\lambda_{j,it})} (y_{j,it} - z)^{\lambda_{j,it}-1} \exp[-\theta_{j,it}(y_{j,it} - z)], \hspace{0.2cm} y_{j,it} \geq z,
\end{equation*} and the p.d.f. of $z$ is given by
\begin{equation*}
p(z|\alpha,\gamma)=\dfrac{\gamma^{\alpha}}{\Gamma(\alpha)} z^{\alpha -1} \exp(-\gamma z), \hspace{0.2cm} z > 0,\alpha >0,\gamma >0.
\end{equation*}
The sampling distribution is given by
\begin{equation*}\label{Eq: multivariate Gamma}
p(\boldsymbol{y}_{it}|\alpha,\gamma,\boldsymbol{\lambda}_{it},\boldsymbol{\theta}_{it}) = \dfrac{\gamma^{\alpha}}{\Gamma(\alpha)} \prod_{j=1}^{m} \dfrac{\theta_{j,it}^{\lambda_{j,it}}}{\Gamma(\lambda_{j,it})} \int_{0}^{B}(y_{j,it} - z)^{\lambda_{j,it} -1} \exp(-\theta_{j,it}(y_{j,it} - z)) z^{\alpha - 1} \exp(-\gamma z) \, dz,
\end{equation*}
where $B=\min\{{y_{j,it}:i=1,\dots,n; j=1,\dots,m; t=1\dots,T\}}$, $\boldsymbol{\lambda}_{it} = (\lambda_{1,it},\dots,\lambda_{m,it})^\prime$, $\boldsymbol{\theta}_{it}=(\theta_{1,it},\dots,\theta_{m,it})^\prime$ and $\boldsymbol{\Theta}_{it} = (\boldsymbol{\lambda}_{it},\boldsymbol{\theta}_{it})^{\prime}$ and $\alpha > 0$,$\gamma>0$.  
The second-level observational equation for $i=1,\dots,n$, $j=1,\dots,m$ and $t=1,\dots,T$  is given by
\begin{equation*}
\begin{aligned}
        \log(\boldsymbol{\Theta}_{j,it}) = \boldsymbol{D}^{\prime}_{j,it}\boldsymbol{\delta}_{j,it} + \boldsymbol{S}^{\prime}_{j,it}\boldsymbol{\eta}_{j,it}, 
\end{aligned}
\end{equation*}
where $\boldsymbol{D}_{j,it}=\left(D_{j,it,1}, \dots, D_{j,it,a_{j}}\right)^{\prime}$ is an $a_{j}$-dimensional vector of exogenous predictors with stock-time varying (dynamic) coefficients $\boldsymbol{\delta}_{j, i t}=$ $\left(\delta_{j, i t, 1}, \cdots, \delta_{j, i t, a_{j}}\right)^{\prime}$ and $\boldsymbol{S}_{j, i t}=\left(S_{j, i t, 1}, \cdots, S_{j, i t, b_{j}}\right)^{\prime}$ is a $b_{j}$-dimensional vector of exogenous predictors with static coefficients $\boldsymbol{\eta}_{j}=\left(\eta_{j, 1}, \cdots, \eta_{j, i t, b_{j}}\right)^{\prime}$. Let $\boldsymbol{\xi}_{i t}$ be a $p_{d}$-dimensional vector constructed by stacking the $a_{j}$ coefficients $\boldsymbol{\delta}_{j, i t}$ for $j=1, \cdots, m$. The structural equation of the HMDM relates the stock-time varying parameter $\boldsymbol{\xi}_{i t}$ to an aggregate (pooled) state parameter $\boldsymbol{x}_{t}$ :
\begin{equation*}
\boldsymbol{\xi}_{i t}=\boldsymbol{x}_{t}+\boldsymbol{v}_{i t}    
\end{equation*}
where the errors $\boldsymbol{v}_{i t}$ are assumed to be i.i.d. $N_{p_{d}}\left(\mathbf{0}, \boldsymbol{V}_{i}\right)$. The state (or system) equation of the HMDM is:
\begin{equation*}
\boldsymbol{x}_{t}=\boldsymbol{G} \boldsymbol{x}_{t-1}+\boldsymbol{w}_{t}
\end{equation*}
where $\boldsymbol{G}$ is a $p_{d} \times p_{d}$ state transition matrix and the state errors $\boldsymbol{w}_{t}$ are assumed to be i.i.d. $N_{p_{d}}(\mathbf{0}, \boldsymbol{W})$. This model could be explored in the context of analyzing the joint dynamics of stock indexes. Although we have not discussed about incorporating dependence between the components of $\boldsymbol{Y}$ through copulas, this could be an interesting topic for further exploration.

\section*{Appendix}\label{appendix}

\subsection*{\texttt{rgeneric()} construction for VAR(1)}
In this section we briefly explain how to build an \texttt{rmodel} to fit latent VAR(1) using \texttt{rgeneric()} in \texttt{R-INLA}. We have to define \texttt{rmodel} that we label \texttt{`inla.rgeneric.VAR1.model'}. Through the \texttt{inla.rgeneric.define()} function,
we create an \texttt{inla.rgeneric} object that we call \texttt{model.VAR}.
\begin{center}
\texttt{model.VAR <- inla.rgeneric.define(inla.rgeneric.VAR1.model, \dots)}.    
\end{center}
Then, we can embed this ``inla.rgeneric" object in the usual INLA syntax to fit latent VAR(1).
\begin{center}
\texttt{formula.inla <- $ \dots + \text{f(idx, model = model.VAR, \dots)} +$ \dots},   \end{center}
where \texttt{idx} refers to the indices in \texttt{R-INLA} setup.

Let $\boldsymbol{x}_{t}$ be a $m$-dimensional time series following  VAR($1$) process as in equation (\ref{Eq: VAR1}). Assuming $\boldsymbol{W} = \text{diag}(\sigma^{2}_{w1}, \dots, \sigma^{2}_{wm})$, be the $m \times m$ covariance matrix of the latent states. The set of parameters in latent VAR($1$) consists of $\{\phi_{ij}, i=1\dots,m; j=1,\dots,m\}$ and the marginal variances $\sigma^2_{wj}$, for $j=1,\dots,m$. 
For the internal representation of the parameters in INLA, we transform the marginal variances into log-precisions such that $\boldsymbol{\theta} = (\phi_{11},\phi_{21},\dots,\phi_{mm},\log(1/\sigma^2_{w1}),\dots,\log(1/\sigma^2_{wm}))^{\prime}$ such that each of its components has $\mathbb{R}$ as its support, as required by \texttt{R-INLA}. 
For more details about how to build an \texttt{rgeneric()} function, \texttt{vignette("rgeneric", package= "INLA")} line in \texttt{R} provides documentation.

In order to define latent VAR$(1)$, we need to construct \texttt{`inla.rgeneric.VAR1.model'}, which returns several functions such as  \texttt{Q()}, \texttt{graph()}, \texttt{mu()}, \texttt{log.prior()}, \texttt{initial()},  \texttt{log.norm.const()}, etc. 
\begin{itemize}
    \item The \texttt{Q()} function defines the precision matrix of the VAR$(1)$ process which is derived in (\ref{Eq: VAR1_precision}).
    
    \item The \texttt{graph()} function defines the 0/1 representation of the precision matrix indicating the zero and non-zero entries of the precision matrix as returned by \texttt{Q()}.
    
    \item The \texttt{mu()} function returns $\boldsymbol{0}$.
    
    \item  We assign a Gaussian prior $\mathcal{N}(\text{mean} = \mu,\text{variance}=\sigma^2)$ on $\phi_{ij}$, for $i=1,\dots,m$, $j=1,\dots,m$ and Gamma prior $\Gamma(.,a;b)$ (with mean $a/b$ and variance $a/b^2$) on the precisions, $1/\sigma^2_{wj}$, $j=1,\dots,m$. Hence the joint prior for $\boldsymbol{\theta}$ becomes 
    \begin{equation}\label{Eq: rgeneric_prior}
        \pi(\boldsymbol{\theta}) =  \prod_{i=1}^{m^2}\mathcal{N}(\theta_{i},1;1) \times  \prod_{j = m^2 +1}^{m^2+m} \Gamma(\theta_{j};1,1) \times \exp(\theta_{j}),
    \end{equation} 
    where for each $j = m^2 + 1,\dots,m^2 + m$, $\exp(\theta_{j})$ is the Jacobian for the change of variable from $1/\sigma^2_{j}$ to $\theta_j= \log(1/\sigma^2_{wj})$.
    Hence the \texttt{log.prior()} returns the logarithm of the joint prior as in (\ref{Eq: rgeneric_prior}).

\item We have defined the initial values of $\phi_{ij} = 0.1$ and precisions $1/\sigma^{2}_{wj}$ as $1$ for $i=1,\dots,m$ and $j=1,\dots,m$. The function \texttt{initial()} contains the initial values of the parameter $\boldsymbol{\theta}$ in the internal scale.

\item  The \texttt{log.norm.const()} denotes the logarithm of the normalizing constant from a multivariate Gaussian distribution.
\end{itemize}

\newpage
\bibliographystyle{abbrvnat} 
\bibliography{biblio-HFT-survey}

\end{document}